\documentclass{PoS}

\title{Towards precise relativistic b quarks on the lattice}

\ShortTitle{Towards precise relativistic b quarks on the lattice}

\author{C. McNeile \\
University of Glasgow, Glasgow, UK \\
E-mail: \email{c.mcneile@physics.gla.ac.uk}}

\author{ C.T.H. Davies \\
University of Glasgow, Glasgow, UK \\
E-mail: \email{c.davies@physics.gla.ac.uk}}

\author{\speaker{E. Follana}\\%
        Universidad de Zaragoza, Zaragoza, Espa\~na\\
        E-mail: \email{efollana@unizar.es}}

\author{K. Hornbostel \\
Southern Methodist University, Dallas, TX, USA \\
E-mail: \email{kjh@physics.smu.edu}

}
\author {G.P. Lepage \\
Cornell University, Ithaca, NY, USA \\
E-mail: \email{g.p.lepage@cornell.edu}}

\author{J. Shigemitsu \\
The Ohio State University, Columbus, Ohio, USA \\
E-mail: \email{shige@pacific.mps.ohio-state.edu}}

\author{HPQCD collaboration \thanks {http://www.physics.gla.ac.uk/HPQCD}}

\abstract{We discuss the status of our ongoing efforts to improve on
  our calculation of the $D_s$ decay constant. We show preliminary
  results on the ratio of the charm to the strange quark mass. We also
  present preliminary results for spectroscopy, decay constants and
  bottom quark mass obtained by performing calculations with highly
  improved staggered quarks at masses above the c mass and close to
  the b mass.}

\FullConference{The XXVII International Symposium on Lattice Field Theory\\
                 July 26-31, 2009\\
                 Peking University, Beijing, China}

\begin{document}

\section{Introduction}

In the last few years Lattice QCD has become a precision tool, useful
for the calculation of non-perturbative, gold-plated quantities in
hadronic physics. This allows us to make stringent tests of QCD, and
is crucial if we are to use Lattice QCD calculations as an ingredient
in the search for physics beyond the Standard Model. It also makes it
possible to determine accurately, from first principles, many
parameters of the Standard Model (strong coupling constant, quark
masses, elements of the CKM matrix.)

Recently it has become possible to include charm quarks accurately in
Lattice QCD calculations, thanks to the introduction of the Highly
Improved Staggered Quark (HISQ) action \cite{charm}. This is a fully
relativistic discretization, and has a number of advantages, which we
discuss later on, as well as some potential difficulties due to the
much higher mass of c quarks as compared with u, d and s quarks. Our
results for charm-light decay constants \cite{fds}, as well as for the
charm quark mass \cite{mc} show that this approach can produce results
with a similar accuracy to the previous staggered light quark
calculations \cite{HPQCD_MILC}.

Having seen that it's possible to use this approach with charm quarks,
the question immediately comes to mind of whether the same is true for
bottom quarks, with ensembles already existing or that can be
realistically generated in the near future. In this exploratory study,
we increase the mass of the HISQ quarks beyond the charm mass into the
bottom mass region.

We briefly discuss general issues with relativistic discretizations of
heavy quarks in section \ref{relat}. We then present in section
\ref{fDs} an update on our calculation of charm-light decay constants,
and section \ref{McMs} presents the status of a calculation of the
ratio $m_c/m_s$, which when combined with our previous accurate value
for $m_c$ will result in an improved $m_s$ determination. In section
\ref{b} we discuss preliminary results for masses and decay constants
of bottom-bottom and bottom-strange systems, obtained through the
relativistic discretization at masses beyond the charm quark mass and
an extrapolation in the quark mass to get to the b. The same strategy
is applied to the calculation of the b quark mass in section
\ref{mb}. Section \ref{conclusions} presents our conclusions and
outlook for future work.

\section{Relativistic Discretization of Heavy Quarks}
\label{relat}

The use of the ASQTAD improved staggered discretization in Lattice QCD
has made it possible, in recent years, to fully include the effect of
u, d and s quarks on the QCD vacuum, resulting in realistic and
accurate calculations of many gold-plated properties of hadrons
\cite{HPQCD_MILC}. The ASQTAD discretization removes all tree-level $a^2$
errors present in the naive staggered action. Recently we introduced a
new, further improved staggered action called HISQ (highly improved
staggered quarks). The HISQ action goes beyond $a^2$ tree-level
improvement by further smearing the gauge fields which enter in the
staggered Dirac operator. We showed in \cite{charm} that this
effectively cuts down the size of the 1-loop taste-changing
interactions, which were typically larger than expected in the
staggered formulation, as well as the ones liable to cause theoretical
concerns if not under control \cite{Sharpe}.

A relativistic discretization has a number of advantages with respect
to an effective heavy-quark theory such as NRQCD. If the discretized
action has some symmetries, there are quantities which are free from
renormalization. For example, for staggered quarks, pseudoscalar meson
decay constants do not renormalize because of PCAC. This eliminates
one of the main sources of systematic errors from the
calculation. Using the same formulation for the heavy and the light
quarks is simpler, and allows us, for example, to calculate accurate
ratios of quark masses.

But using a relativistic action for heavy quarks brings forth new
issues. Discretization errors will be generally proportional to powers
of the quark mass, $(am)^n$, and the question is whether those can be
kept under control for large masses. For staggered actions only even
powers appear, and for an action like HISQ that removes all $a^2$
tree-level errors, the discretization errors start at $(am)^4$,
$\alpha_s (am)^2$. These would still be too large for precision work;
however, the fact that a system with charm (or heavier) quarks is
essentially non-relativistic makes it possible, by a simple
modification of the coefficient of the Naik term, to completely remove
those errors at leading order in $v/c$, where $v$ is the velocity of
the quark inside the hadron \cite{charm}. Another essential ingredient
is the existence of fine enough ASQTAD sea quark ensembles
\cite{asqtad_MILC}, going from $a \approx $ 0.15 fm (very coarse),
0.125 fm (coarse), 0.09 fm (fine), 0.06 fm (superfine) and down to
0.045 fm (ultrafine). All these ingredients make it possible to use
the HISQ action for calculations with charm quarks with almost the
same level of precision as was possible for light quarks. The
existence of ensembles at several lattice spacings is crucial to
reliably extrapolate to the continuum limit.  For bottom quarks, at
present we must also extrapolate in the heavy quark mass: we do
calculations at several values of the lattice spacing and the heavy
quark mass (above the charm mass), and make a joint $a \to 0$, $M_h
\to M_b$ extrapolation. On the ultrafine ensemble such extrapolation
is small.

\section{$f_{D_s}$ decay constant}
\label{fDs}

\begin{figure}[h]
  \begin{center}
    \begin{tabular}{cc}
      \resizebox{80mm}{!}
                {\includegraphics{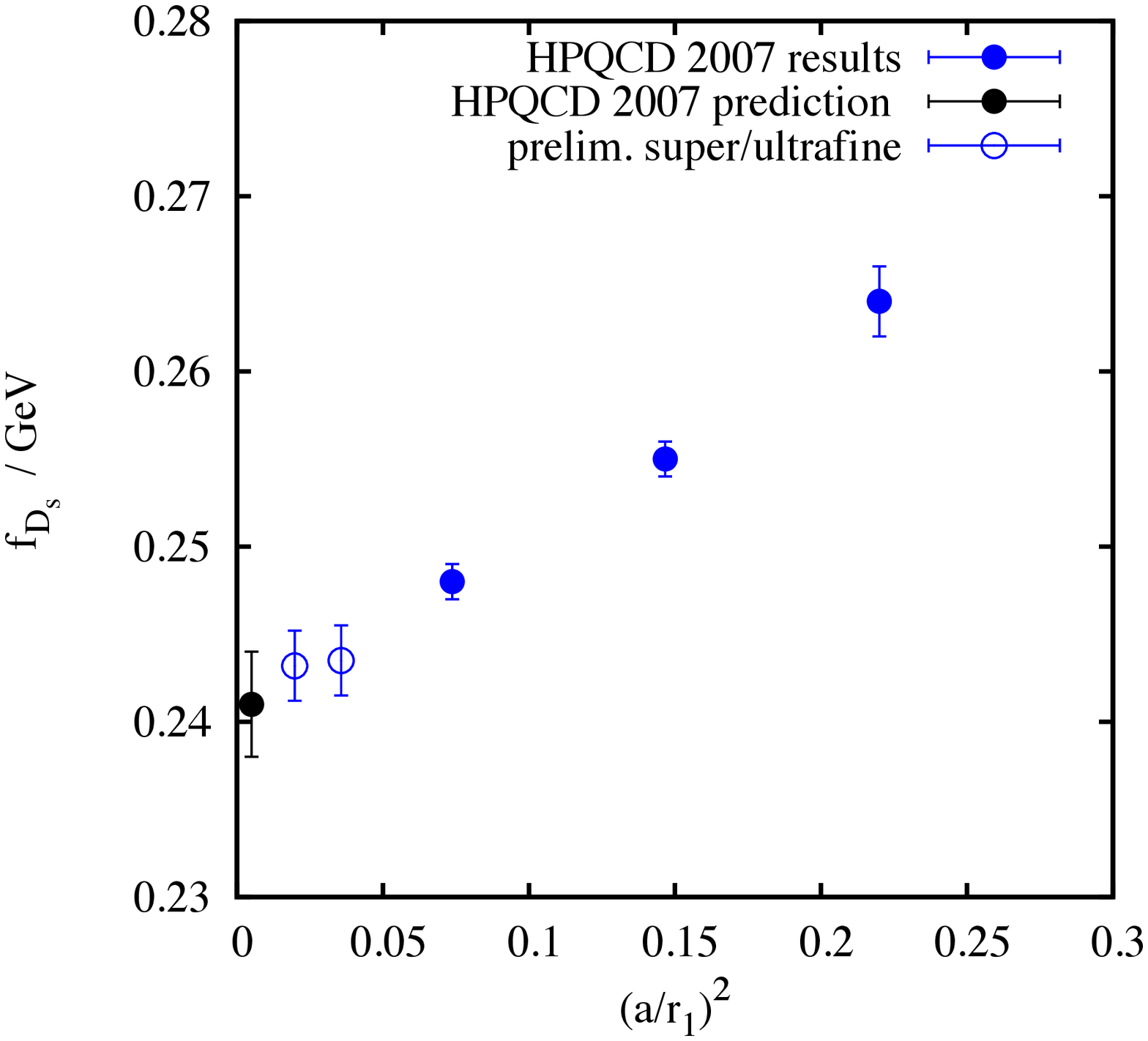}} &
      \resizebox{!}{70mm}
      {\includegraphics{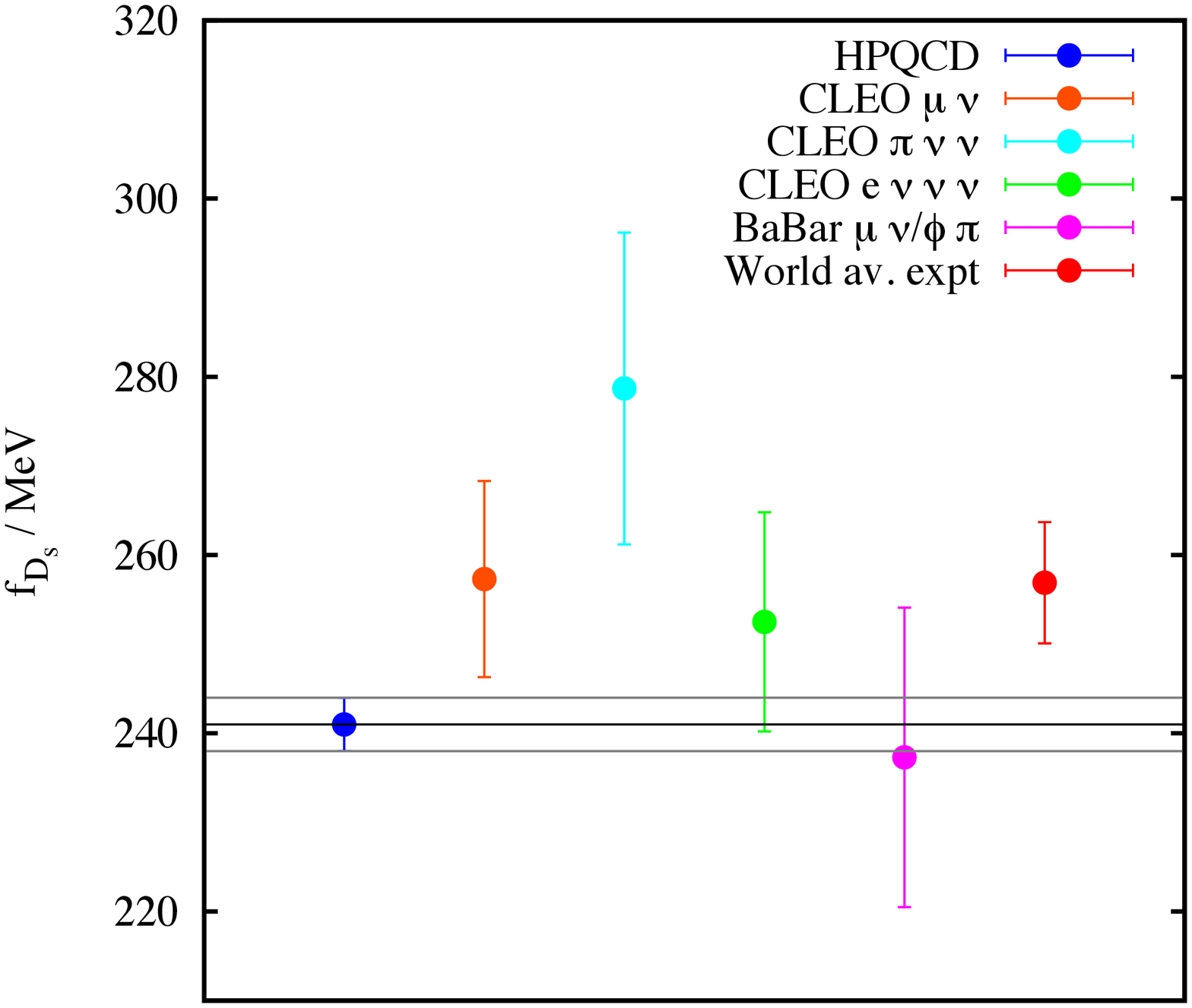}} \\
                   \\
    \end{tabular}
    \caption{HPQCD lattice calculation and comparison with
      experiment. World average is from HFAG \cite{HFAG}}
    \label{fds}
  \end{center}
\end{figure}

Our prediction of the $D$ and $D_s$ decay constants caused a lot of
interest because experimental determinations soon after \cite{CLEO-c1,
  CLEO-c2, CLEO-c3} gave a surprising picture: $f_D$ agreed very well
with our result but $f_{D_s}$ was more than $3 \sigma$ higher
\cite{gamiz, lat08}. Here $\sigma$ was dominated by the experimental
error since our own error was so small.  Because we use a relativistic
discretization in which there is a PCAC relation, we can calculate
$f_D$ and $f_{D_s}$ from the partially conserved axial current, in
exactly the same way as is done for $f_\pi$ and $f_K$ (for which we
agree with the experimental values). Using three values of the lattice
spacing, we obtained a $2\%$ accurate result. We now have extended the
calculation to two finer lattice spacings, the superfine $a \approx
0.06$ fm, and the ultrafine $a \approx 0.045$ fm. We show in
Fig. \ref{fds} the preliminary results from this updated calculation,
and a comparison of our data with current experimental determinations
which have changed the picture considerably from that of last
year. Now new CLEO determinations \cite{CLEO-c4, CLEO-c5} are closer
to our result for $f_{D_S}$ (especially for some channels) and a
reanalysis of BaBar data \cite{BaBar} is very close indeed. The
discrepancy between our calculation and the current world average
given by the Heavy Flavor Averaging Group is only about $2 \sigma$
\cite{HFAG}. \footnote{In fact the final discrepancy should be even a
  little smaller, because our final result will use the updated
  determination of the parameter r1 \cite{r1}, which is used to set
  the scale and therefore the physical value of the lattice spacing.
  r1 has decreased a little from its previous calculated value, and
  that in turn will increase slightly the value of $f_{D_s}$.}.

\section{Ratio of $m_c$ to $m_s$}
\label{McMs}

Because we are using the same HISQ action for all quarks, light and
heavy, we can obtain very accurate ratios of quark masses, where some
systematic errors cancel. In particular, we have a very precise value
for the ratio of lattice masses, $m_c/m_s$. We can now go to another
scheme, like $\overline {MS}$, and the Z factors that appear will
cancel in the ratio; this allows us to combine this ratio with our
precision calculation of the $m_c$ mass \cite{mc} to obtain a
determination of the strange quark mass with a $\approx 1.5\%$ total
error.
               
\begin{figure}[h]
\begin{center}
\includegraphics[width=.6\textwidth]{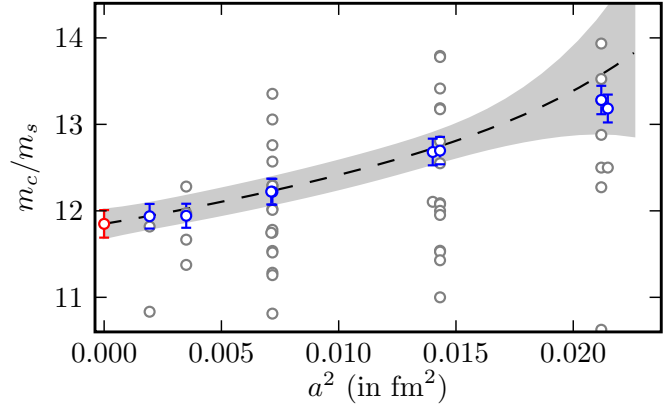}
\caption{Ratio of charm to strange quark masses.}
\label{mcms}
\end{center}
\end{figure}

In Fig. \ref{mcms} we show preliminary results for $m_c/m_s$. The
gray points in the plot are the lattice results for the ratio at
various bare masses; the blue points are the results interpolated to
the physical masses, determined by fixing the masses of the $\eta_c$
and the $\eta_s$ to their physical values\footnote{See \cite{r1} for a
  discussion on what we mean by the ``physical'' value of $\eta_s$;
  the main point is that it can be used to accurately fix the strange
  quark mass.}; the red point is the continuum and chirally
extrapolated value.

\section{Masses and decay constants for bottom-bottom and
  bottom-strange systems}
\label{b}

In order to study bottom quarks with current lattices, we must resort
to an extrapolation procedure: we calculate at several values of the
lattice spacing $a$ and masses above the charm mass, $M_h$, and then
we do a joint continuum and mass extrapolation, $a \to 0$, $M_h \to
M_b$. 

\begin{figure}[h]
\begin{center}
\includegraphics[width=.99\textwidth]{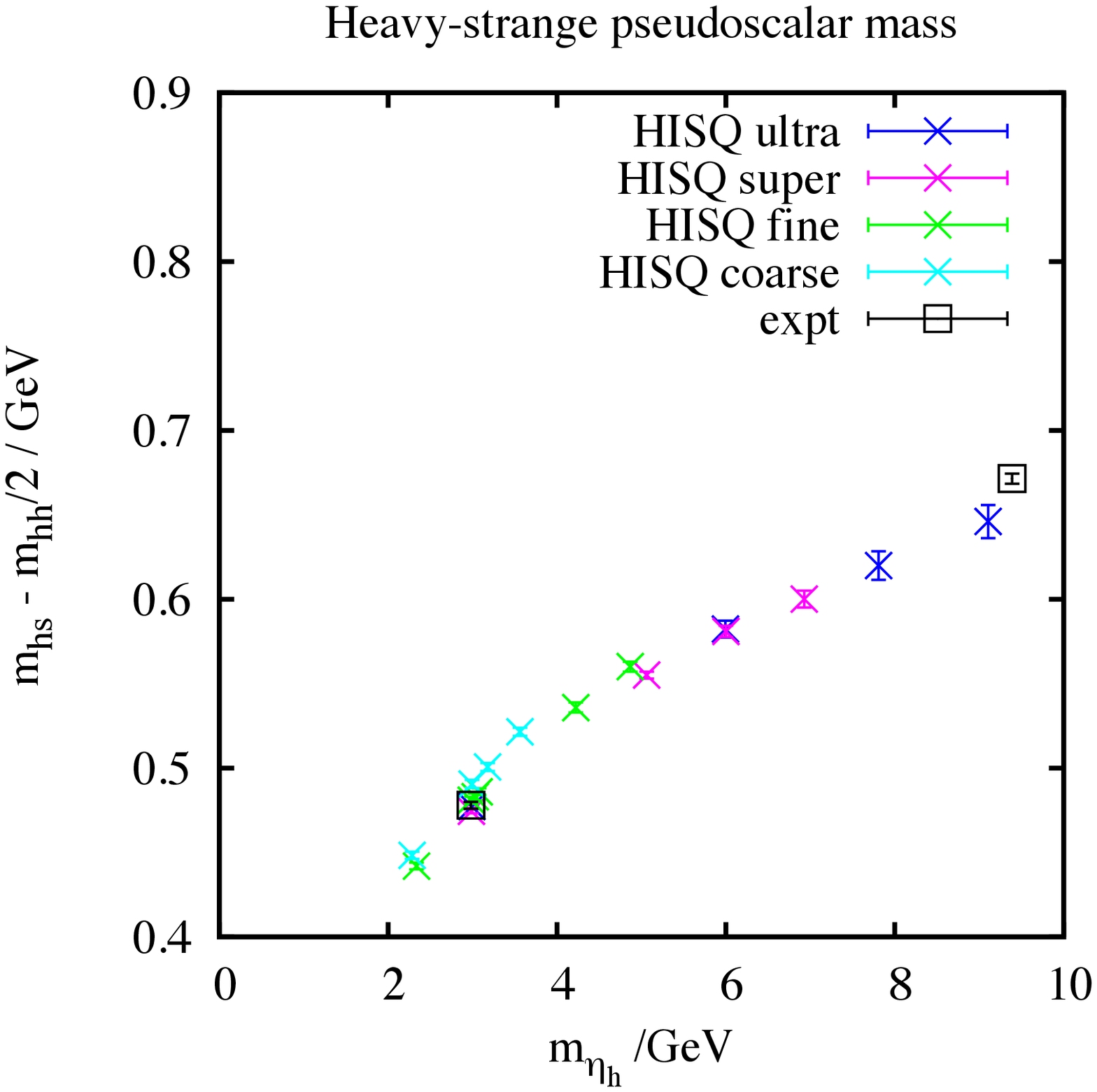}
\caption{$\Delta M_{hs} = M_{hs} - M_{hh} / 2$}
\label{mh}
\end{center}
\end{figure}

In Fig. \ref{mh} we show preliminary results for the difference
between the masses of the heavy-strange and the heavy-heavy
pseudoscalar mesons, $\Delta M_{hs} = M_{hs} - M_{hh} / 2$, for a
range of masses from the charm to the bottom quark mass and for
several lattice spacings. 
We can see that the results have a smooth dependence on both the heavy
mass and the lattice spacing, and the extrapolated value to the bottom
mass will be consistent with experiment.

\begin{figure}[h]
\begin{center}
\includegraphics[width=.99\textwidth]{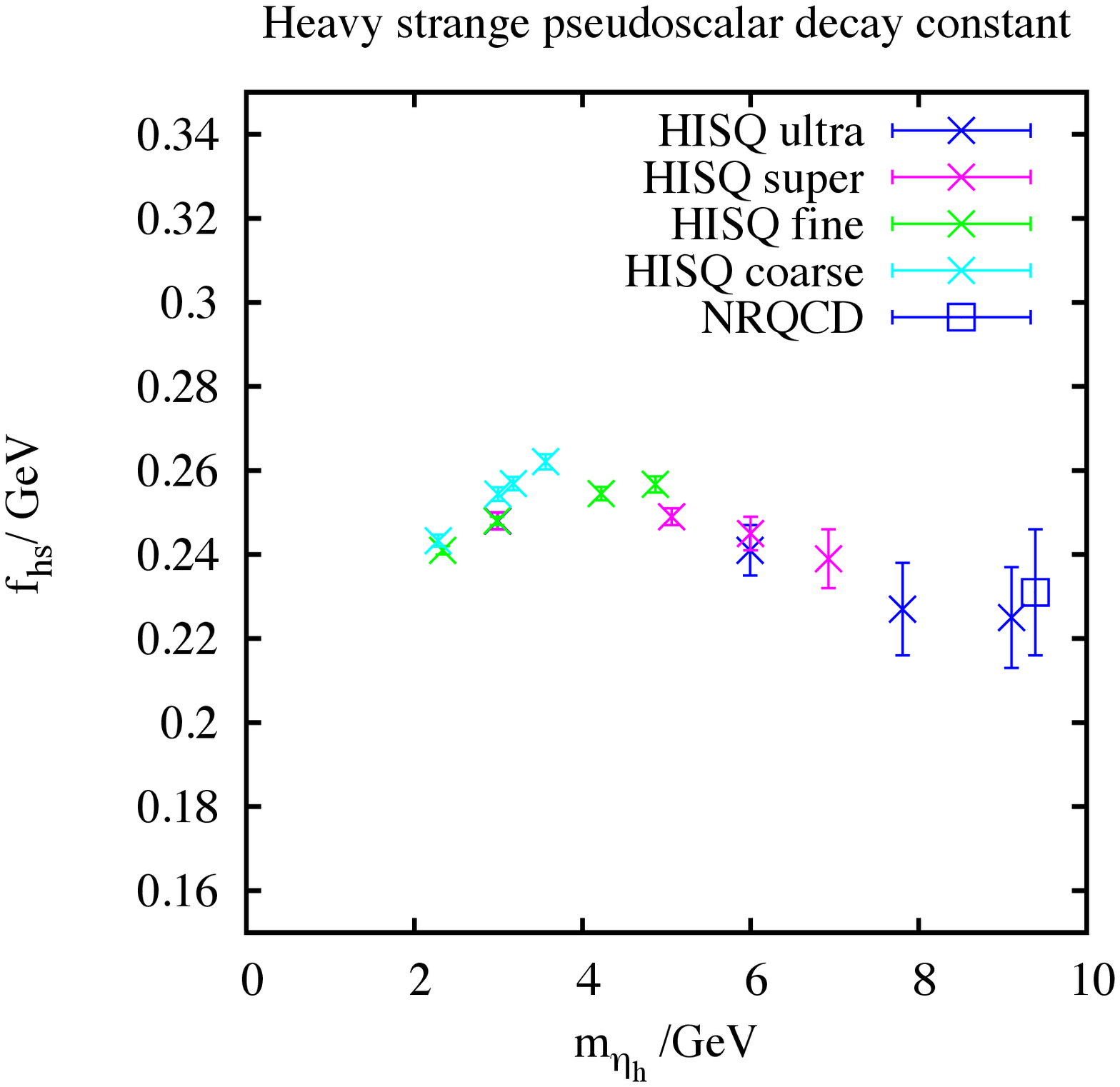}
\caption{$f_{hs}$}
\label{fh}
\end{center}
\end{figure}

In Fig. \ref{fh} we show similar data for the decay constant of the
heavy-strange pseudoscalar, $f_{hs}$, which interpolates between
$f_{D_s}$ and $f_{Bs}$. 
The extrapolated value should be consistent, and already competitive,
with the previous NRQCD calculation \cite{NRQCD_fbs}.

\section{Bottom quark mass}
\label{mb}

Another thing which we can do with the relativistic formulation is a
calculation of the bottom quark mass, with the same methods used for
the charm quark mass calculation \cite{mc}. Let's recall that the
basic idea is to take advantage of the high-order continuum QCD
perturbation calculations performed by Chetyrkin, K\"{u}hn,
Steinhauser and Sturm, by combining them with high precision lattice
data for moments of pseudoscalar and vector current correlators. For
the case of a pseudoscalar current, the basic object is the
correlator:
\begin{equation}
G(t) \equiv a^6 \sum_{\bf{x}} \left( a{m_0}_h\right)^2 
\left<0|j_5({\bf x}, t) j_5(0,0)|0\right>, \;
j_5 = \bar{\psi}_h\gamma_5\psi_h
\end{equation}
which is finite and unrenormalized as $a \to 0$, due to PCAC. The
quantities to compare with continuum PT are the moments, $G_n = \sum_t
(t/a)^n G(t)$, of such correlator, for which we have:
\begin{equation}
G_n(a=0) = \frac{g_n(\alpha_{\overline{MS}}(\mu),
  \mu/M_h)}{(aM_h(\mu))^{n-4}}
\end{equation}
Here the $g_n$ are known to large orders in continuum PT for large
$M_h$ and not too large $n$, and the left hand side is calculated from
the lattice data, giving us the value of $a M_h$. In practice we use
what are called reduced moments, in order to remove some of the
discretization errors (see \cite{mc} for details.)

\begin{figure}[h]
\begin{center}
\includegraphics[width=.6\textwidth]{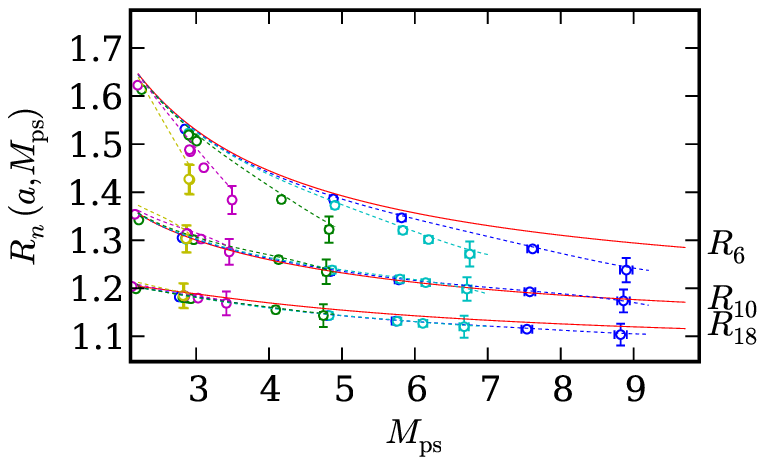}
\caption{6th, 10th and 18th reduced moments of the pseudoscalar
  current-current correlator.}
\label{rn}
\end{center}
\end{figure}

In Fig. \ref{rn} we can see several reduced moments, for different
values of the heavy-heavy pseudoscalar mass and several values of the
lattice spacing, as well as the continuum extrapolations. With this
method we expect the final value on $m_b$ to have a total error of
$2-3\%$.

\section{Conclusions and Outlook}
\label{conclusions}

It is feasible to use a highly improved relativistic action for charm
quarks on fine enough lattices. This allows us to do very accurate
calculations for systems with charm quarks. We obtain a $2\%$ accurate
result for $f_{D_s}$. Due to changes in the experimental values, our
calculation is now only about $2 \sigma$ away from experiment. We also
can obtain a very accurate $m_c/m_s$ mass ratio, which combined with
our $1\%$ $m_c$ determination will provide a $\approx 1.5\%$
determination of $m_s$.

To get to the bottom quark mass, we need (at least for now) to
extrapolate in the mass of the heavy quark. This can be done, as we
show for $\Delta M_{bs} = M_{bs} - M_{bb} / 2$ and $f_{B_s}$, and it
seems that the additional source of error that it introduces can be
kept under control. The results are in both cases consistent with
either experiment or previous NRQCD calculations.

If we had even finer lattices (0.03 fm, say) we could reduce
substantially that uncertainty. This may be a good strategy to reduce
the errors on some B physics quantities, for example $f_{B_s}$ .

\section{Acknowledgments}

We are grateful to the MILC collaboration for the use of their
configurations. Computing was done at USQCD's Fermilab cluster, the
Ohio Supercomputer Center and the Argonne Leadership Computing
Facility at Argonne National Laboratory, which is supported by the
Office of Science of the U.S. Department of Energy under contract
DOE-AC02- 06CH11357. We acknowledge the use of Chroma \cite{chroma}
for part of our analysis. This work was supported by the Leverhulme
Trust, NSF, the Royal Society, the Scottish Universities Physics
Alliance, STFC and DOE. E.~Follana is supported by Ministerio de
Ciencia e Innovaci\'on through the Ram\'on y Cajal program.

\end{document}